# Non thermal isostructural electrically driven insulator-metal transition and electro-strain in layered cobaltate

Abdul Ahad[1], R. Rawat[2], F. Rahman[1], D. K. Shukla[2,*]

[1]Department of Physics, Aligarh Muslim University, Aligarh 202002, India
[2]UGC-DAE Consortium for Scientific Research, Indore 452001, India

*Correspondence to: dkshukla@csr.res.in

## Abstract:

We report here a discovery of electrically driven insulator to metal transition (IMT) and concomitant isostructural volume expansion in the layered cobaltates which otherwise do not exhibit temperature dependent IMT. These findings are demonstrated at macroscopic, microscopic and atomic scales. With application of voltage growth of metallic regions have been observed in the 2D layered $La_{2-x}Sr_xCoO_4$. Growth of metallic regions is associated with volume increase (strain as high as 0.3%). Non thermal IMT and electro-strain are proposed to be caused by electro-proliferation of the $Co^{3+}$ high spin states.

## Introduction:

Materials ($VO_2$ & $NbO_2$) which exhibit electrically driven insulator to metal transition (IMT) have attracted huge interests, as these offer potential applications such as, memristors[1,2], sensors[3], neuromorphic computing[4] and transistor switches[5]. In these materials electric field induced resistive state change happens due to existing temperature dependent IMT, through the Joule heating[6,7]. Moreover, IMT in these materials is accompanied by structural transitions. To find a material which can exhibit isostructural non thermal IMT is a challenge and desirable for fast application purposes[6].

Compounds containing $Co^{3+}$ ions usually show complex electrical and magnetic behavior and unpredictable phase diagram because of intriguing spin-state transitions. The Co ions in +3 valence state can exhibit three different spin-states i.e. low spin state (LSS) ($t_{2g}^6 e_g^0$), intermediate spin state (ISS) ($t_{2g}^5 e_g^1$) and high spin state (HSS) ($t_{2g}^4 e_g^2$). Each of this spin-states has different ionic radii and electronic structure. These three spin states are almost degenerate and easily can be changed into other by small external perturbations i.e. temperature, pressure and magnetic field[8,9]. As a result of spin state transition, lattice compresses/expands and physical properties changes like conductivity etc. One of the perturbations that can also change the spin-states is the electric field. However, there is rarely any report on electric field tuning of these.

Layered cobaltates $La_{2-x}Sr_xCoO_4$ are insulators and exhibit mix of the LSS and HSS of the $Co^{3+}$ ions at room temperature[10]. These do not exhibit temperature dependent insulator to metal transition. We present here observations on electrically driven IMT based on proliferation of the $Co^{3+}$ HSS in these compounds. Along with IMT due to electro-proliferation of the HSS concomitant electro-strain is also observed. Combination of both these electrically driven non-thermal phase transitions can be extremely useful for future multisensory and memristor applications.

## Results:

Electrically driven resistive switching and simultaneous observation of strain in the x=0 composition of the $La_{2-x}Sr_xCoO_4$ (Fig. 1 (a) and Fig. 2 (a & b)) is surprising and very interesting. Simultaneous resistance and strain switching observation motivates to compare results with composition having different amount of the $Co^{3+}$ fraction. Other than x = 0 (100% $Co^{3+}$) we choose x = 0.5 (50 % $Co^{3+}$ and 50 % $Co^{2+}$ ($Co^{2+}$ always remain in the HSS)) composition. We have performed strategic set of measurements on both these samples. Hereafter, these samples will be termed as S1 (x = 0) and S2 (x = 0.5).

Resistivity of the sample S1 is much lower than the sample S2 and both samples show insulating character. With increasing temperature resistance of both the samples decreases. For temperature dependent measurements we have chosen separate range of temperatures where resistances are similar for both the samples. Resistance vs temperature (RT) of both samples with different applied voltages are shown in the Fig. 1 (a & c). Both samples show voltage induced resistance switching. Due to current compliance limit (10 mA) of measuring device, at switching all the curves show constant value of resistance and temperature coefficient of resistance (TCR) of switched state remains unknown. To figure out the TCR we have employed another device with high current measurement ability. A clear metallic nature (+ve TCR) after the switching is observed (see Fig. 1 (b)). This confirms voltage induced IMT in the sample. This measurement has been done in heating cycle because of high current flowing in metallic phase prevents cooling.

It is well proven that strain gauge is very sensitive to small changes in the adjacent systems[11]. Fig. 2 (a) and 2 (b) show IV and strain measurements for the sample S1 in the temperature range 190 K to 300 K and Fig. 2 (c) and 2 (d) comprises same measurements for the sample S2 in the temperature range 320 K to 400 K. A remarkable novel association between lattice expansion and IV is observed. Interestingly, to one's imagination RT curves shows breakdown with the different applied voltages but IV clearly shows that the phase transition is reversible and behavior is like a threshold type resistive switching[12] and clearly not a breakdown[13–15] as observed in Mott insulators and charge ordered systems.

Behind the observed electrically driven IMT and electro-strain there could be two reasons, first there is a tunneling between the $Co^{3+}$ HSS, and second proliferation of $Co^{3+}$ HSS. Confirmation to either of these is not possible without microscopy (conducting microscopy) at submicron length scale. Therefore, we have performed conducting tip force microscopy (CTFM) measurement on the sample S2 to know the mechanism at micro scale. In this microscopy voltage bias is applied between the tip and sample. Changes in local resistance is measured with the applied bias. CTF microscopy can provide ~ 10 nm spatial resolution[16]. We have first recorded topography of the sample at 4 V and selected 4 x 2 $\mu m^2$ area to scan the current flowing via tip for different bias voltages. Fig. 3 clearly show growth of the conducting regions with increasing applied bias. This microscopic evidence supports our macroscopic measurements.

To further check the atomic scale involvement in volume change at unit cell level we have performed X-ray diffraction (XRD) with applied voltages. Fig. 4 shows the XRD measurements in the Bragg-Brentano geometry on the sample S2, for higher 2θ range as it offers sensitivity to changes at smaller inter-planar spacings. Interestingly, we observe successive shifting of diffraction peaks towards lower 2θ with the applied voltage indicating the unit cell expansion. This further supports to our finding at atomic scale. We do not find any evidence of structural change as is generally observed in the transition metal dichalcogenides[17].

## Discussion:

Coming back to the strain value which is associated with metallic region growth and switching is the novel observation. These spectacular changes can be mapped with normal strain gauge (as the changes are sufficiently large). To track the path of phase formation we have employed the three geometries of voltage biasing and simultaneous measurement of current and strain, parallel, perpendicular and sidewise. In parallel geometry strain gauge is mounted back to sample along the area and voltage is applied on the opposite face. In perpendicular geometry strain gauge is mounted on one face and voltage is applied across the thickness. In sidewise geometry, voltage is applied on the same face (on the side) at which strain gauge mounted. We have presented these measurements for the S2 (see Fig. S1). The noteworthy observation is that the no spectacular change in the strain of the sample is observed in the sidewise case (see Fig. S2). Moreover, the maximum changes occur in the parallel case. This observation clearly indicates the growth of metallic regions depends on the electric field gradient.

In the last decades, several reports[2,6,7,13–15] have shown IMT type switching but the origin of those are not very clear. However, the controversies regarding such phases have evidences of joule heating i.e. the phase conversion is actually filamentary type and a consequence of the Joule heating[18]. Recently, in the layered $Ca_2RuO_4$, it has been observed that the electrically induced metallic phase is not a result of Joule heating and the structure is involved in this case but the details are not known[7]. In the present case, we can discard the involvement of the Joule heating based on the arguments that voltage induced IMT and volume expansion directly depends on the geometry (as explained above).

In the following we discuss a new viewpoint that the host to the strain in these samples is the $Co^{3+}$ HSS. The ionic radii of $Co^{3+}$ HSS is the 0.61 Å and for the LSS it is 0.545 Å[19]. Considering the unit cell comprised of 100 % HSS or 100 % LSS then there should be volume difference of ~ 6 %. S1 has 100 % $Co^{3+}$ and S2 has 50 % $Co^{3+}$, therefore, if the applied voltage in these compounds changes the spin states of the $Co^{3+}$ to complete HSS then the strain should be ~ 50 % larger in S1. To confirm that the IMT and electro-strain is due to the HSS transition we have compared the values of strain in the S1 and S2 at temperatures (190 K for S1 and 330 K for S2) where they have similar resistance. Strain values of S1 (at 190 K) is ~ 40% larger than S2 (at 330 K), a difference as predicted is observed. A self-explanatory schematic to this process is shown in the Fig. 5.

## Conclusions:

In conclusion, we have provided the first experimental evidence of electric field induced IMT, strain and spin state transition in the 2D layered $La_{2-x}Sr_xCoO_4$ (x= 0, 0.5). Volume expansion and simultaneous growth of metallic regions is due to the electric field induced spin-state transition. Electric field driven growth of metallic region at submicron scale is confirmed by microscopy (CTF) and expansion of lattice by macroscopic (strain gauge) and atomic scale (X-ray diffraction) measurements. Unprecedented phenomenon observed in the present study could be useful for future advance devices.

## Acknowledgments:

Authors are grateful to Ajit Sinha (Director, UGC-DAE CSR) for encouragement and support. Authors are thankful to D. M. Phase and G. Panchal for help during XRD measurements. AA acknowledges UGC, New Delhi, India for financial support in the form of MANF scheme. DKS acknowledges financial support from Department of Science and Technology (DST), New Delhi, India.

## Author contribution:

DKS conceived the idea of the project and devised the experiments. AA made the samples with the help from FR. AA and DKS developed the required software performed the experiments, and analyzed the data. RR performed the microscopy experiment and participated in data analysis. DKS wrote the paper with the feedback from all the authors.

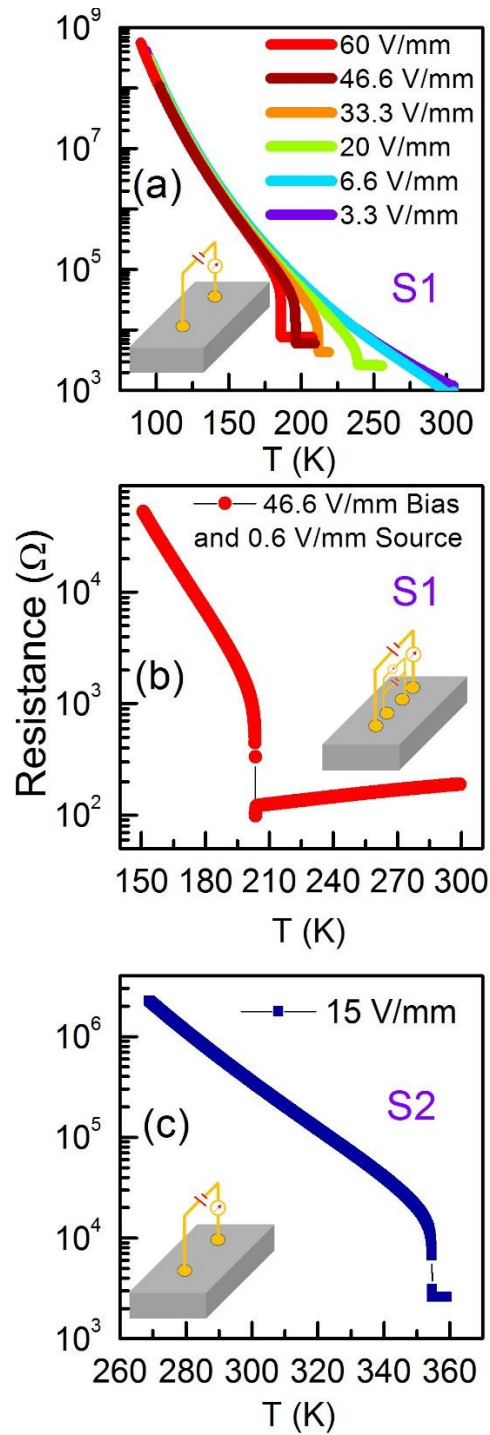

Fig.1. (a) Temperature vs resistance (RT) measurement of the sample S1 under different electric fields. (b) RT of the sample S1 under 46.6 V/mm electric field at outer leads, and current measurements through inner leads using high compliance limit instrument (see text for details). During current measurement through inner leads a small electric field of 0.6 V/mm is applied. (c) RT of the sample S2 under 15 V/mm electric field. Measurement configurations are schematically shown in the insets of respective figures.

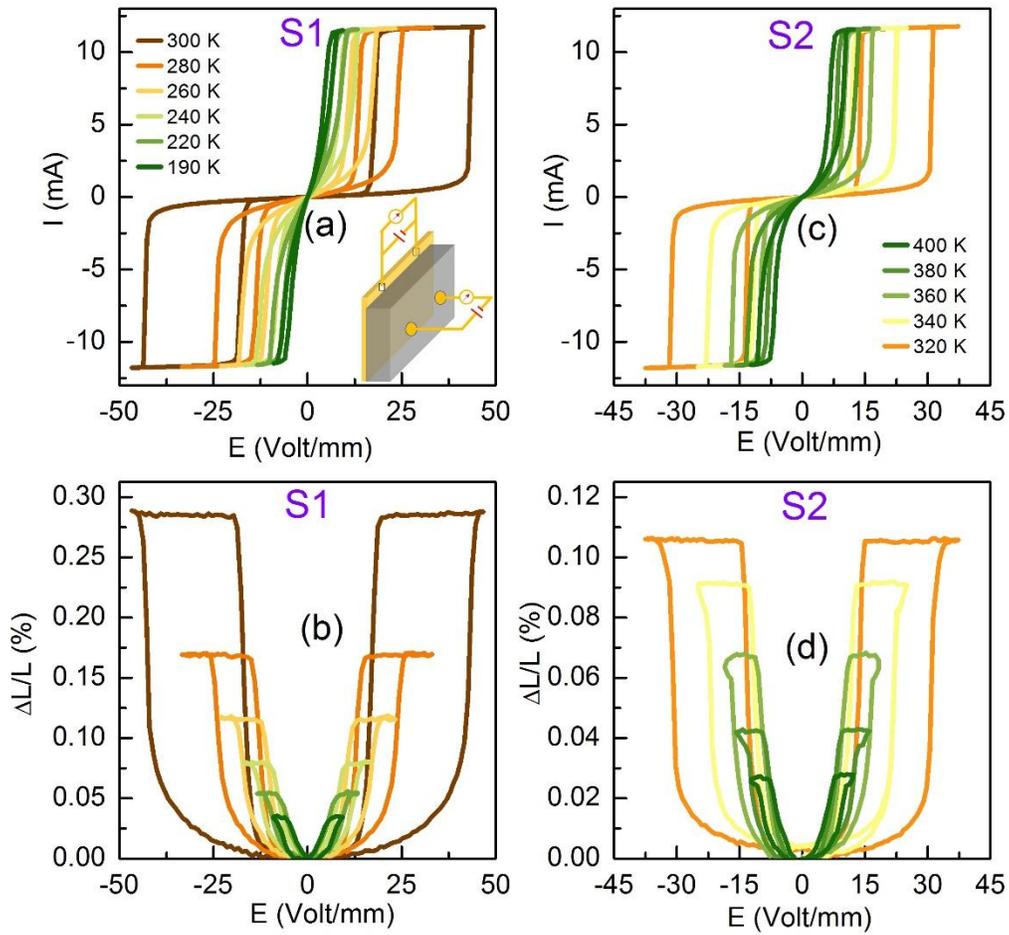

Fig. 2. (a) Current vs voltage (IV) measurements of the sample S1 at different temperatures (190 K-300 K) and (b) strain measurements. (c) IV measurements of the sample S2 at different temperatures (320 K-400 K), and (d) strain measurement. For both the samples (S1 and S2) IV and strain measurements were performed simultaneously, IV using two-probe and resistance of strain gauge using four-probe, as shown in the schematic in the inset of the figure (a).

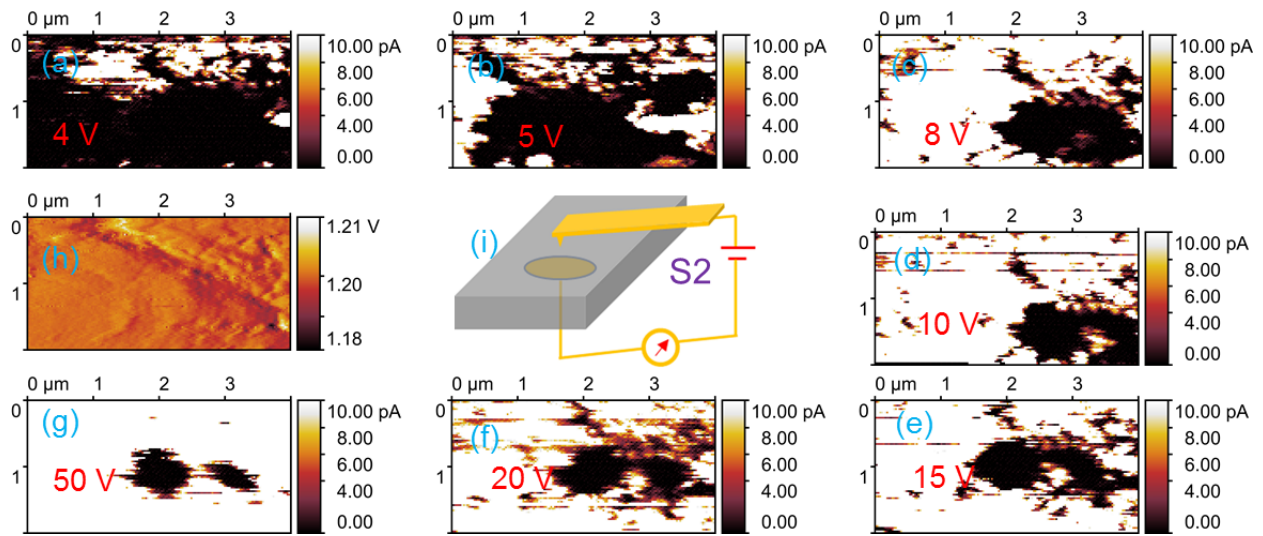

Fig. 3. (a-g) Current scanning images of the sample S2 under labelled biasing voltage. Each figure is direct evidence to the proliferation of metallic regions. (h) Topography of the sample S2 (i) Schematic of the CTFM measurement configuration.

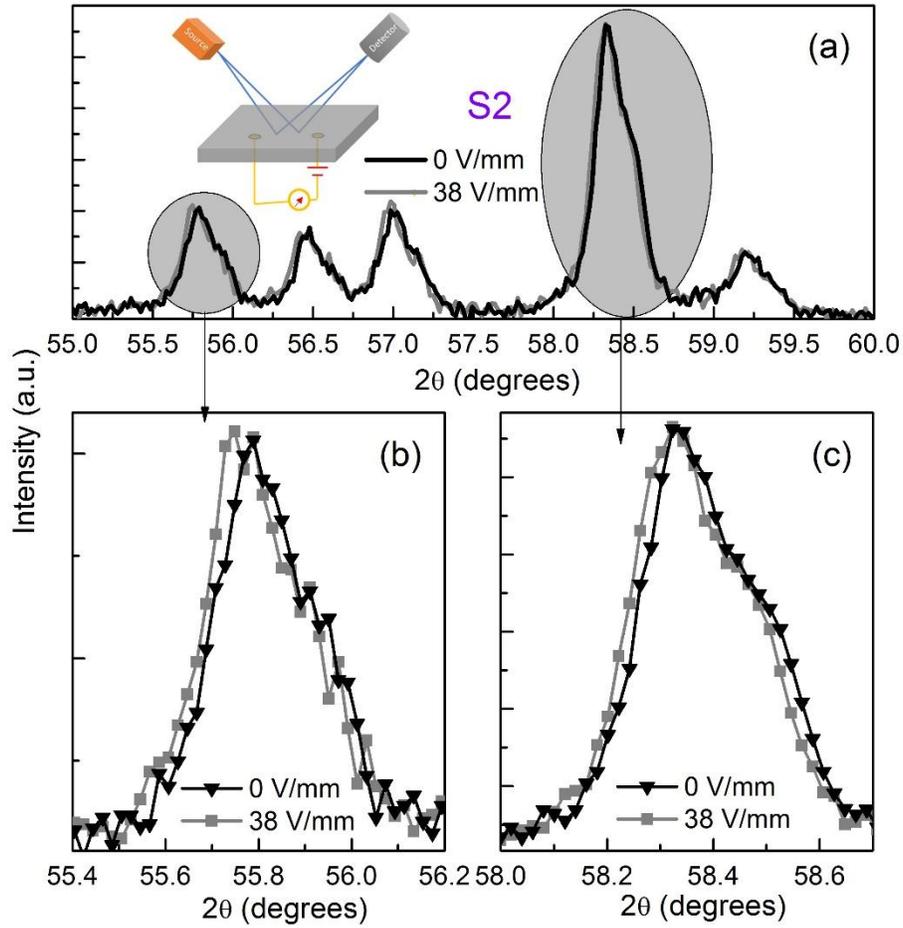

Fig. 4. (a) XRD of the sample S2 with and without electric field. Inset shows the experimental configuration of XRD under electric field, and (b & c) show the zoomed views of the selected diffraction peaks.

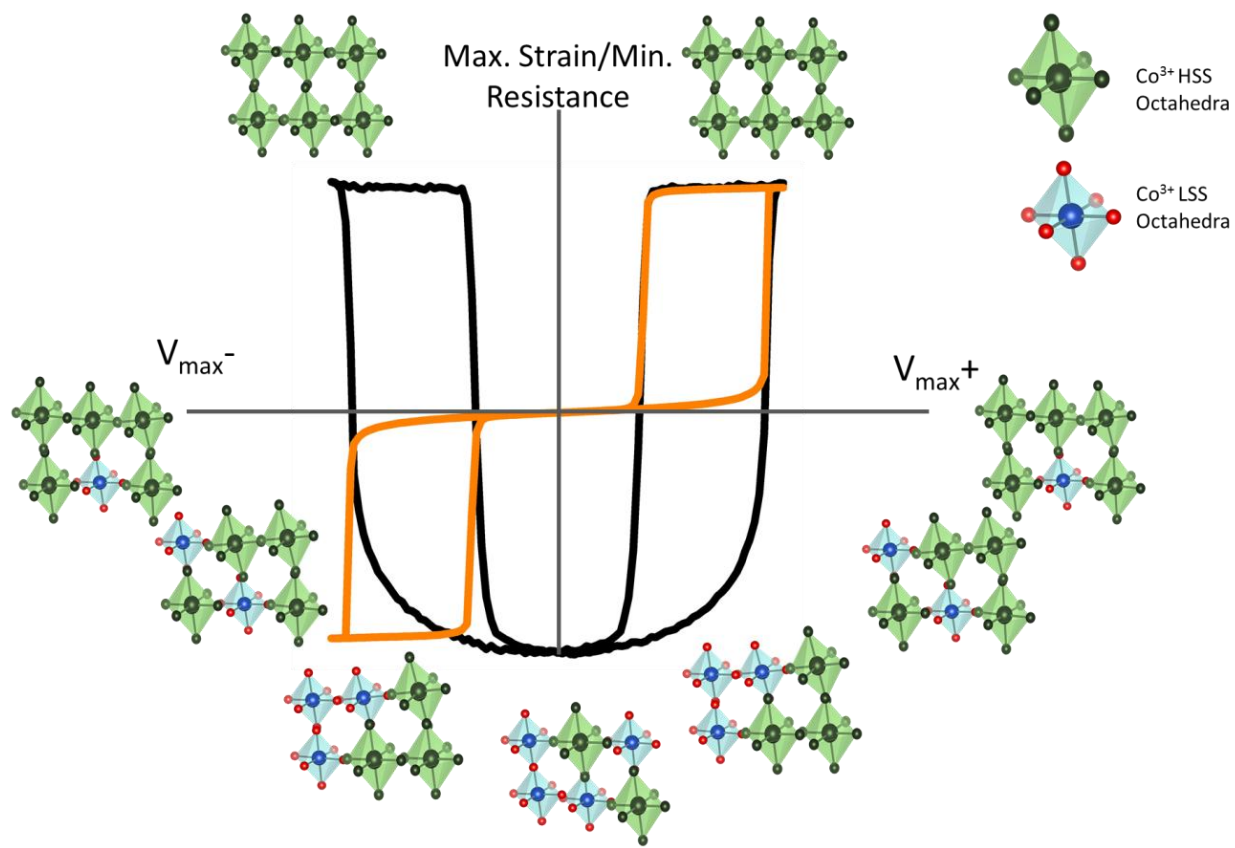

Fig. 5. Schematic of the electro-proliferation of the $Co^{3+}$ HSS with the application of applied electric field and simultaneous strain generation/IV behavior.

# Supplementary information

# Non thermal isostructural electrically driven insulator-metal transition and electro-strain in layered cobaltate

Abdul Ahad[1], R. Rawat[2], F. Rahman[1], D. K. Shukla[2,*]

[1]Department of Physics, Aligarh Muslim University, Aligarh 202002, India
[2]UGC-DAE Consortium for Scientific Research, Indore 452001, India

*Correspondence to: dkshukla@csr.res.in

## Materials and Methods:

Polycrystalline samples of $La_{2-x}Sr_xCoO_4$ (x = 0.0 and 0.5) were synthesized using standard solid-state synthesis method and are well characterized [10,20]. For current-voltage characteristic (IV) two probe as well as four probe method has been employed. Keithley 2182A nanovoltmeter, Keithley 2401 sourcemeter and Keithley 6517B electrometer were used for the electrical measurements in different configurations. Silver point contacts have been made for each electrical measurements. Applied voltage (Volts) has been converted into electric field (V/mm), normalizing by the distance of the contacts. To perform the strain measurements, strain gauge (Tokyo Sokki Kenkyujo Co. Ltd.) was glued on sample surface by GE varnish. Simultaneous resistance measurements were performed for sample (two probe using electrometer) and strain gauge resistance (four probe). Room temperature conducting tip force microscopy (CTFM) was done on attocube made commercial system. During CTFM the biasing up to 10 V has been used by inbuilt voltage source of setup and for higher bias voltages Keithley 6517B was used. Room temperature X-ray diffraction patterns have been recorded on a Bruker diffractometer in Bragg-Brentano geometry with a custom made sample holder for XRD voltage biasing.

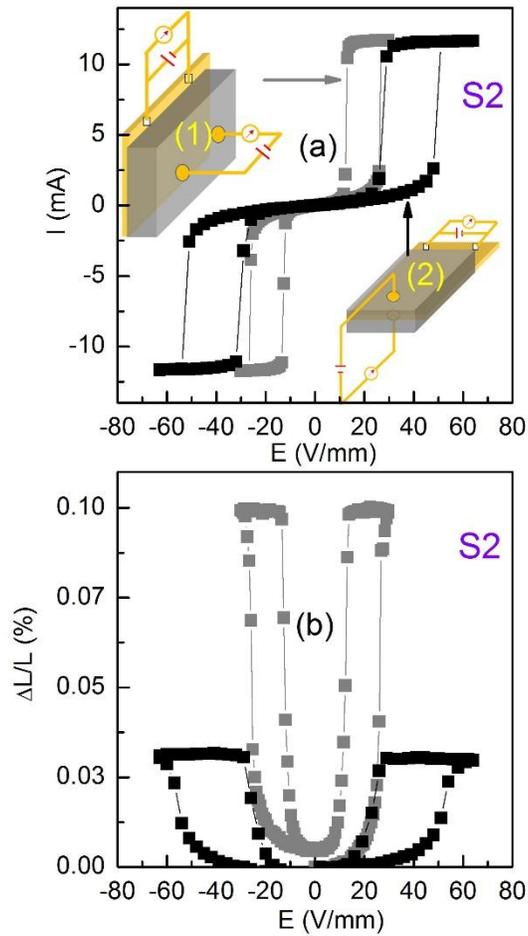

Fig. S1. (a) IV characteristic for the sample S2 measured at 330 K and (b) associated strain measurement. In the parellel geometry (1) lesser electric field is required for IMT switching and larger strain is observed (grey curves), while in the perpendicular geometry (2) larger electric field is required for IMT switching and lesser strain is observed (black curves). This experiment confirms that the expansion is caused due to the electric field gradient. The geometry (1) is much sensitive for changes because the gradient is applied on the whole sample piece and strain is measured across the gradient. The geometry (1) is commonly used throughout this study.

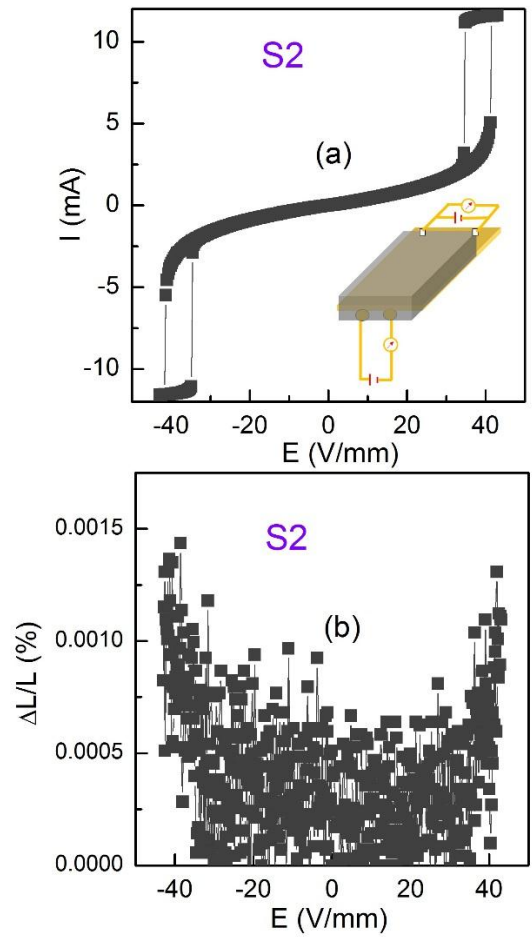

Fig. S2 (a) IV characteristic for the sample S2 measured at 350 K and (b) associated strain measurement. In this measurement configuration (side wise geometry, shown in inset of (a)) no appreciable change is observed with applied electric field.